\documentclass[
aps,
pra,
reprint,
superscriptaddress,
longbibliography
]{revtex4-1}

\usepackage{graphicx}
\usepackage{dcolumn}
\usepackage{bm}

\begin{document}

\title{Increased cyclicity of atomic transitions via coherent interference of decay paths}

\newcommand{\bra}[1] {
  \langle #1 |}
\newcommand{\ket}[1] {
  | #1 \rangle}

\newcommand{\tab}[0]{\ \ \ \ \ \ \ \ \ \ \ \ \ \ }
  
\author{Eliza Cornell}
\affiliation{John A. Paulson School for Engineering and Applied Sciences, Harvard University, Cambridge, Massachusetts, USA}

\author{Benjamin Pingault}%
\affiliation{Pritzker School of Molecular Engineering and Chicago Quantum Institute, University of Chicago, Chicago, Illinois, USA}
\affiliation{Q-NEXT and Materials Science Division, Argonne National Laboratory, Lemont, Illinois, USA}

\author{Gergő Thiering}%
\affiliation{HUN-REN Wigner Research Centre for Physics, P.O.\ Box 49, H-1525 Budapest, Hungary}

\author{Neil Sinclair}
\affiliation{John A. Paulson School for Engineering and Applied Sciences, Harvard University, Cambridge, Massachusetts, USA}

\author{Ádám Gali}
\affiliation{HUN-REN Wigner Research Centre for Physics, P.O.\ Box 49, H-1525 Budapest, Hungary}
\affiliation{Department of Atomic Physics, Institute of Physics, Budapest University of Technology and Economics,  M\H{u}egyetem rakpart 3., H-1111 Budapest, Hungary}
\affiliation{MTA–WFK Lend\"{u}let ``Momentum" Semiconductor Nanostructures Research Group, P.O.\ Box 49, H-1525 Budapest, Hungary}

\author{Marko Lončar}
\affiliation{John A. Paulson School for Engineering and Applied Sciences, Harvard University, Cambridge, Massachusetts, USA}

\date{\today}

\begin{abstract}
Optical readout is a fundamental tool in atomic state measurement, yet the fidelity of optical readout is frequently limited by imperfect photon collection. This can be mitigated when readout occurs on a cycling transition which continuously fluoresces under resonant excitation, thus increasing signal and enabling single-shot readout. We present a method to extend the cyclicity of atomic transitions via coherent destructive interference between spurious decay paths. We describe the characteristics of atomic systems in which this method can be implemented and model several examples in which the number of emitted photons is increased by multiple orders of magnitude.

\end{abstract}

\maketitle

\section{Introduction}

Optical cycling transitions are an important feature of atomic systems, allowing high-fidelity state measurement in solid-state quantum defects \cite{rosenthal_single-shot_2024,robledo_high-fidelity_2011,raha_optical_2020,sukachev_silicon-vacancy_2017}  and trapped ions and atoms \cite{monroe_programmable_2021, bernien_probing_2017, morgado_quantum_2021}. Cycling transitions are additionally used to cool and trap atoms and ions \cite{phillips_laser_1982, raab_trapping_1987, leibrandt_cavity_2009,shuman_laser_2010}. Deviation from ideal cyclicity due to spurious decay channels decreases the fidelity of state readout \cite{banks_resonant_2019,rosenthal_single-shot_2024}, and reduces the efficiency of cooling and trapping. The pursuit of improved cyclicity in optical transitions has lead to recent advances in molecular design \cite{klos_prospects_2020,dickerson_fully_2022,mitra_pathway_2022,dragan_features_2023}. Researchers have also demonstrated selective cavity enhancement of optical transitions which increases the system’s effective cyclicity \cite{raha_optical_2020,gritsch_optical_2025}. This relies, however, on a cavity linewidth which is narrower than the spectral separation between the desired and the undesired decay paths, and in some atomic systems this cavity design may be technologically infeasible. This article discusses another means of increasing cyclicity, the coherent interference of decay paths controlled via externally applied driving fields.

The phenomenon of coherent interference of decay has been proposed for tailoring spectral lineshapes \cite{zhu_spectral_1996,ficek_simulating_2004}, population inversion for lasing purposes \cite{lee_quenching_1997,mandel_inversionless_1993}, and squeezing \cite{ficek_simulating_2004}. It is closely related to coherent population trapping \cite{berman_analysis_1998}, which is usually framed in terms of a dark superposition state which cannot be coherently excited by a specific applied field. Interference of decay, on the other hand, involves a dark superposition state which does not spontaneously decay.

This article focuses on the relative enhancement and suppression of different decay paths for the purpose of increasing the cyclicity of a given transition, rather than suppressing all decay, as has been emphasized in past investigations. We investigate two approaches. The first relies on the specific form of the atomic system’s transition dipole moments (TDMs) to create an excited state superposition under a single strong drive. For this approach, we will give an example of a specific quantum emitter that is a prime candidate, the negatively charged nickel vacancy center in diamond. The second approach uses an auxiliary field to couple the excited manifold to an auxiliary state, allowing for drive-based control of the superposition. For simplicity, we limit all of our analysis to systems with only two decaying excited states, but the principles we investigate here can be generalized to more complicated excited state manifolds.

\section{Cycling transitions induced by a strong drive} \label{sec:strongDrive}

\begin{figure*}
    \centering
    \includegraphics{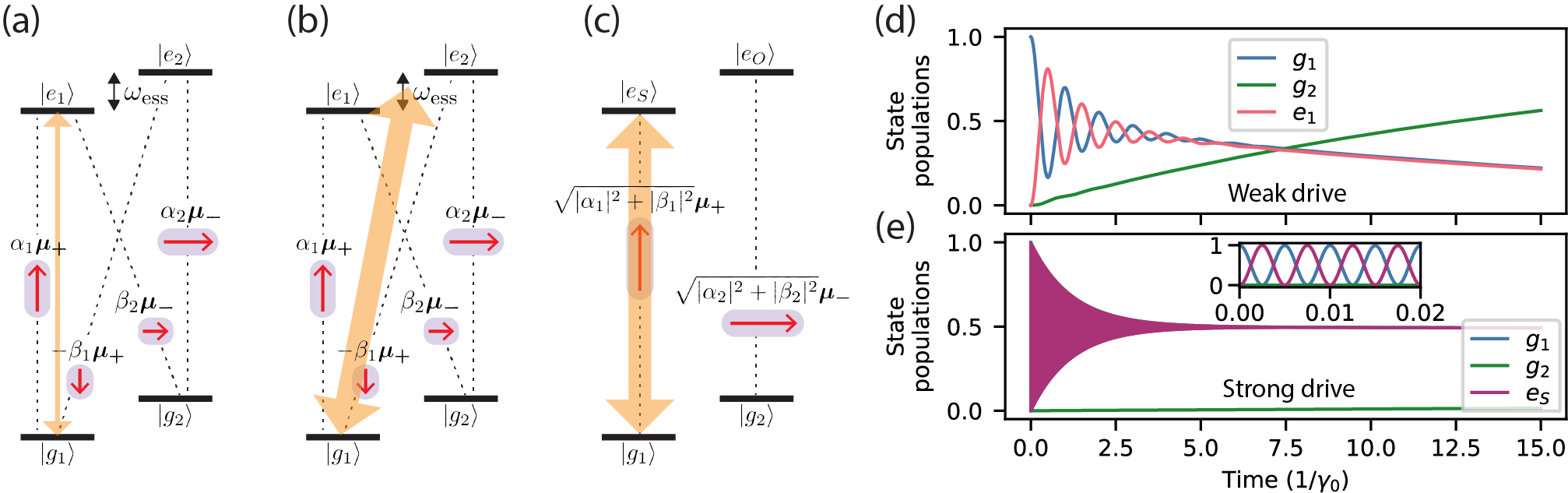}

    \caption{Cycling transitions via strong drive. \textbf{(a)} 
    An atomic system without native cycling transitions. In a traditional readout scheme that solely couples $\ket{g_1}$ to $\ket{e_1}$, the number of photons emitted will be limited because the system will be optically pumped to $\ket{g_2}$. \textbf{(b)} The same atomic system with a strong drive polarized along $\boldsymbol{\mu_+}$ which simultaneously couples $\ket{g_1}$ to both $\ket{e_1}$ and $\ket{e_2}$. When the Rabi rate is much greater than the frequency splitting of the excited states $\omega_\textrm{ess}$, the system is directly driven between $\ket{g_1}$ and $\ket{e_S} \sim \alpha_1 \ket{e_1} - \beta_1 \ket{e_2}$. \textbf{(c)} The system with the excited states basis rewritten, indicated the allowed transitions. In (a-c), the amplitude and polarization of the TDMs are indicated with red arrows. \textbf{(d)} State populations simulated for a weak field coupling $\ket{g_1}$ to $\ket{e_1}$, as in (a), with optical Rabi rate $\Omega = |\mathbf{E}\cdot \boldsymbol{\mu_{e_1g_1}}| = 0.1\omega_\textrm{ess}$. \textbf{(e)} State populations simulated for a strong drive coupling $\ket{g_1}$ to both $\ket{e_1}$ and $\ket{e_2}$, as in (b) and (c), with an effective optical Rabi rate of $\Omega_S = 20\omega_\textrm{ess}$. The coherent behavior at short time scales is shown in the inset. For (d) and (e) the decay from $\ket{e_1}$ to $\ket{g_1}$ has a rate of $\gamma_0 = 0.1\omega_\textrm{ess}/2\pi$ and a branching ratio of b=0.9. We also set $|\alpha_1|=|\alpha_2|$ and $|\beta_1|=|\beta_2|$. 
    }
    \label{fig:strongDriveLevels}
\end{figure*}

Consider an atomic system with ground states $\ket{g_1}$ and $\ket{g_2}$, excited states $\ket{e_1}$ and $\ket{e_2}$, and corresponding TDMs of $\boldsymbol{\mu_{e_i,g_j}} = \bra{e_i}\boldsymbol{d}\ket{g_j} $ . If all four excited-to-ground decay paths are allowed, then this system does not have native cycling transitions. A traditional optical readout scheme to determine whether the system is in $\ket{g_1}$ or $\ket{g_2}$ relies on driving a single optical transition, for example, the $\ket{g_1}-\ket{e_1}$ transition (Fig. \ref{fig:strongDriveLevels}(a)). If the system starts in state $\ket{g_1}$, the average number of emitted photons (and thus the measurement fidelity) is limited by the branching ratio of $\ket{e_1}$ into $\ket{g_1}$, given by:
\begin{equation}
    b_r = \frac{|\boldsymbol{\mu_{e_1g_1}}|^2}{|\boldsymbol{\mu_{e_1g_1}}|^2+|\boldsymbol{\mu_{e_1g_2}}|^2}
\end{equation}
A branching ratio of $b = 1$ describes a perfect cycling transition. When $b < 1$, the system is optically pumped into $\ket{g_2}$.

However, consider the case where the TDMs are of the form: 
\begin{eqnarray}\label{eq:idealTDMs}
\boldsymbol{\mu_{e_1g_1}} &=& \alpha_1\boldsymbol{\mu_{+}} \nonumber\\
\boldsymbol{\mu_{e_2g_1}} &=& -\beta_1 \boldsymbol{\mu_{+}} \nonumber\\
\boldsymbol{\mu_{e_1g_2}} &=& \beta_2\boldsymbol{\mu_{-}} \nonumber\\
\boldsymbol{\mu_{e_2g_2}} &=& \alpha_2 \boldsymbol{\mu_{-}} 
\end{eqnarray}
with $\boldsymbol{\mu_{+}}, \boldsymbol{\mu_{-}}$ as orthogonal polarizations and $\alpha_i, \beta_i$ as complex numbers such that $(\frac{\alpha_1}{\beta_1})^*= \frac{\alpha_2}{\beta_2}$. An artificial cycling transition can be induced in this system by a strong coherent drive polarized parallel to $\boldsymbol{\mu_{+}}$, which couples $\ket{g_1}$ to both $\ket{e_1}$ and $\ket{e_2}$ (Fig. \ref{fig:strongDriveLevels}(b)). In the limit of large drive strength, where the optical Rabi rate of the excitation field is much larger than the frequency splitting $\omega_\textrm{ess}$ between the excited states, the driving field detuning becomes negligible and the system is driven between $\ket{g_1}$ and the superposition state $\ket{e_S} = n (\alpha_1 \ket{e_1} -\beta_1 \ket{e_2})$. Here $n$ is an overall normalization constant. Decay from $\ket{e_S}$ to $\ket{g_2}$ is forbidden (Fig. \ref{fig:strongDriveLevels}(c)). The finite probability of populating the orthogonal superposition state $\ket{e_O} = n (\beta_1^* \ket{e_1} +\alpha_1^* \ket{e_2})$, which does have an allowed decay into $\ket{g_2}$, can be made arbitrarily low by increasing the drive strength.

We simulate the state populations under a traditional optical measurement, in which a weak field couples only $\ket{g_1}$ and $\ket{e_1}$, taking $b=0.9$ (Fig. \ref{fig:strongDriveLevels}(d)). The system starts in $\ket{g_1}$ and is optically pumped into $\ket{g_2}$ on a time scale of  about $10/\gamma_0$, where $\gamma_0$ is the bare state decay rate from $\ket{e_1}$ to $\ket{g_1}$ . About 10 photons are emitted on average before the system goes dark. On the other hand, when a strong drive is applied which couples $\ket{g_1}$ to $\ket{e_S}$, the optical pumping into $\ket{g_2}$ is significantly suppressed (Fig. \ref{fig:strongDriveLevels}(e)). Over the same time scale, the population cycles between $\ket{g_1}$ and $\ket{e_S}$, continuously emitting photons with negligible population transfer to $\ket{g_2}$.

\begin{figure}
    \centering
        \includegraphics[width=1\linewidth]{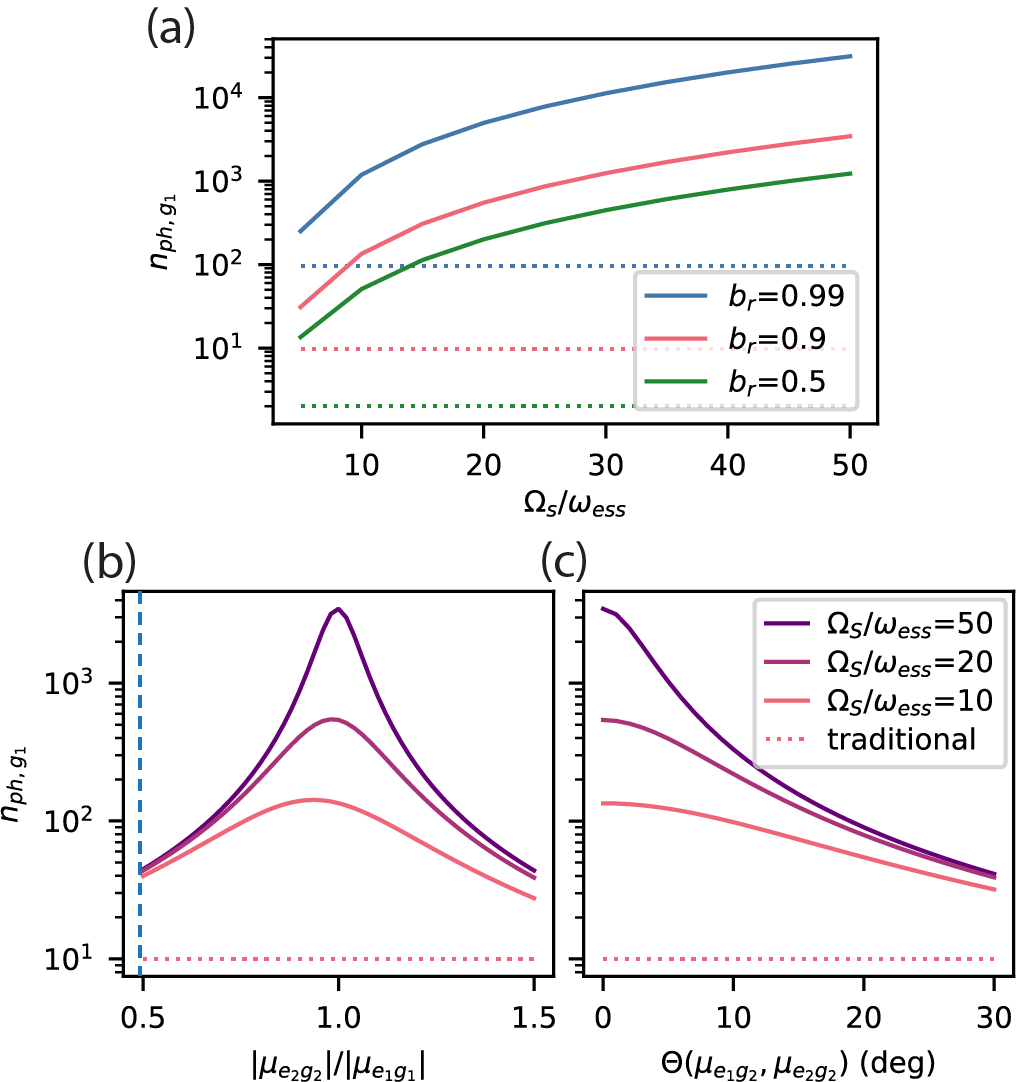}
    \caption{Comparison of strong driving photon emission outcomes for atomic systems with different TDMs. \textbf{(a)} Number of total average emitted photons $n_{ph,g_1}$ from an atomic system whose TDMs have the form in Eq. \ref{eq:idealTDMs}. The dotted lines indicate $n_{ph,g_1}$ for a traditional excitation scheme which applies a weak coherent drive resonant with the $\ket{g_1}-\ket{e_1}$ optical transition. The solid lines indicate $n_{ph,g_1}$ for a system driven by a strong field which is tuned halfway in between the $\ket{g_1}-\ket{e_1}$ and $\ket{g_1}-\ket{e_2}$ transitions. \textbf{(b)} Number of total average emitted photons $n_{ph,g_1}$ for a system in which the TDM amplitudes vary from the form in Eq. 2, parameterized in terms of the ratio between $|\boldsymbol{\mu_{e_2g_2}}|$ and $|\boldsymbol{\mu_{e_1g_1}}|$. \textbf{(c)} Number of emitted photons $n_{ph,g_1}$ for a system in which the TDM polarizations vary from the form in Eq. \ref{eq:idealTDMs}, parameterized by the angle between $\boldsymbol{\mu_{e_1g_2}}$ and $\boldsymbol{\mu_{e_2g_2}}$. In (b) and (c) we observe that for TDMs that vary more significantly from the form in Eq. \ref{eq:idealTDMs}, the system decays more quickly into $\ket{g_2}$ and fewer total photons can be extracted. In (b) and (c) we use $|\boldsymbol{\mu_{e_1g_1}}| = 3 |\boldsymbol{\mu_{e_2g_1}}| = 3 |\boldsymbol{\mu_{e_1g_2}}|$ (giving a branching ratio of $b=0.9$). In (a-c) we  use a bare decay rate of $\gamma_0 = 0.1\omega_\textrm{ess}/2\pi$.
    }
    \label{fig:strongDriveComparison}
\end{figure}

In Fig. \ref{fig:strongDriveComparison} we  simulate $n_{ph,g_1}$, the total average number of photons emitted assuming the system starting in $\ket{g_1}$. The value $n_{ph,g_1}$ is calculated by integrating spontaneous emission events from the master equation (see Appendix \ref{app:Photon counting FOM}). It depends on the system’s original branching ratio $b$ and the superposition optical Rabi rate, given by: \begin{equation}
    \Omega_S = \sqrt{|\mathbf{E}\cdot \boldsymbol{\mu_{e_1g_1}}|^2 + |\mathbf{E}\cdot \boldsymbol{\mu_{e_2g_1}}|^2}
\end{equation}
where $\mathbf{E}$ is the electric field vector of the driving field.

We find that the strong driving technique enhances the number of photons emitted by multiple orders of magnitude over a traditional optical readout scheme (Fig. \ref{fig:strongDriveComparison}(a)). Atomic systems with higher branching ratios $b$ have larger $n_{ph,g_1}$ under traditional excitation, and can be pushed to even higher $n_{ph,g_1}$ values with a strong drive-induced cycling transition.

A lesser increase in expected photon emission is found for systems whose TDMs deviate from the form of Eq. \ref{eq:idealTDMs} either in their amplitude (Fig. \ref{fig:strongDriveComparison}(b)) or their polarization (Fig. \ref{fig:strongDriveComparison}(c)). In this case the expected number of emitted photons saturates as the driving power increases. This is because the decay from $\ket{e_S}$ to $\ket{g_2}$ is non-zero when the TDMs do not fully cancel and becomes the dominant limitation to the cyclicity of the system, restricting it beyond the limitation set by the small probability of excitation into $\ket{e_O}$.

\section{Nickel vacancy center} \label{sec:NiV}
The negatively-charged nickel vacancy center (NiV–) in diamond \cite{thiering_magneto-optical_2021,morris_lifetime-limited_2025} shows promise as a quantum defect, due to its optical stability and long spin coherence time of $>1$ ms at 1.65K  \cite{morris_transition-metal_2026}. The NiV– is a spin-1/2 defect and its excited state manifold comprises only one orbital, so the TDMs of both optical transitions corresponding to a given ground state will have parallel polarizations.

\begin{figure}
    \centering
    \includegraphics[width=1\linewidth]{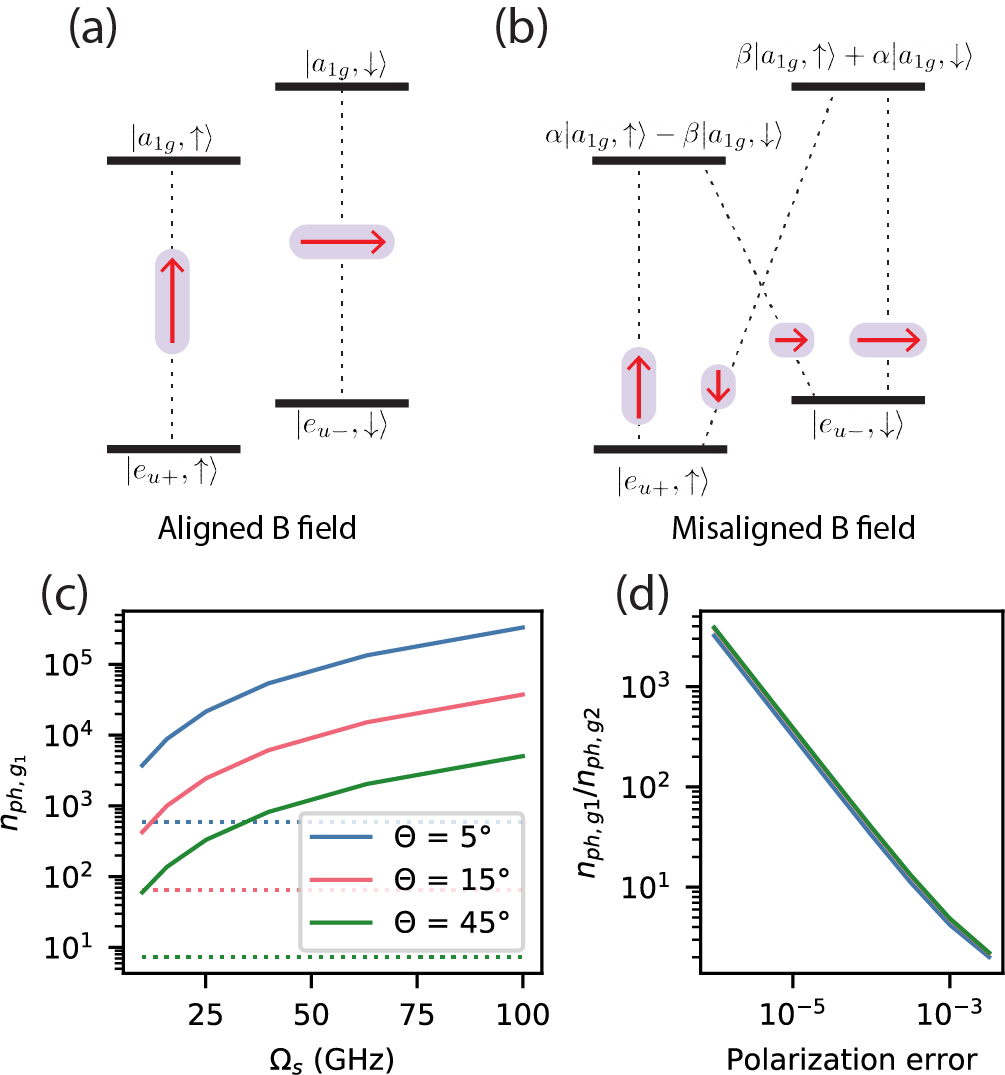}
    \caption{Strong drive-induced cycling transitions in an NiV– center. \textbf{(a)} The NiV–’s two excited states and two lowest-energy ground states under an externally applied magnetic field that is aligned to the symmetry axis of the NiV–. The excited state manifold contains only one orbital, while the two lowest ground states have different orbital characteristics, due to spin-orbit coupling. \textbf{(b)} The NiV– under an externally applied and slightly misaligned magnetic field. The ground states retain the same orbital and spin character, while the excited state magnetic quantization axis rotates freely. All optical transitions are allowed. \textbf{(c)} Number of photons $n_{ph,g_1}$ emitted by the NiV– with a strong drive for readout. The dotted lines indicate results for a traditional weak drive. The simulation uses the optical and spin Hamiltonian given in \cite{thiering_magneto-optical_2021}, an optical decay time of 10ns as measured in \cite{morris_lifetime-limited_2025}, an orbital decay time of 30ps, and a magnetic field amplitude of B = 500 G.  The parameter $\theta$ is the magnetic field angle with respect to the defect axis of symmetry. The splitting between the excited state levels is about 1.5 GHz under these magnetic fields.
    \textbf{(d)} The ratio between $n_{ph,g1}$ and $n_{ph,g2}$ (the number of photons emitted when the system starts in $\ket{g_2}$) simulated over a time of $t=1e3/\gamma_0$ and with optical Rabi rate of $\Omega_S = 100$ GHz, for different polarization errors of the driving field, defined as the fraction of optical power in the undesired polarization. Magnetic field angles of 5°, 15° and 45° result in nearly equivalent results. In order to reduce simulation time, a uniform Purcell enhancement of 100 was assumed for all optical transitions in the strong drive simulations in (c) and (d), increasing $\gamma_0$ to 1/100ps. This enhancement produces a proportional scaling of all decay rates and a broadening of transition linewidths. In the strong drive regime where $\Omega_S \gg \omega_\textrm{ess}$, it does not affect the final value of $n_{ph,g1}$ or $n_{ph,g1}/n_{ph,g2}$.}
    \label{fig:NiVStrongDrive}
\end{figure}

Under a magnetic field aligned to its symmetry axis, the NiV– exhibits native cycling transitions defined by the strong suppression of the spin-flipping transitions (Fig. \ref{fig:NiVStrongDrive}(a)). For certain applications, however, it may be necessary to apply an off-axis magnetic field, for example to create spin-phonon coupling or to increase the speed of optical pumping for state initialization. An off-axis magnetic field tilts the excited state spin quantization axis to a greater degree than that of the ground state spin, because the spin-orbit coupling ``pins” the quantization axes of the ground state spin, while the excited state spin can freely rotate to align with the external magnetic field. Thus under a slightly misaligned magnetic field, the lowest energy ground states are:
\begin{eqnarray}
    \ket{g_1} &\approx& \ket{e_{u+},\uparrow} \nonumber\\
    \ket{g_2} &\approx& \ket{e_{u-},\downarrow}
\end{eqnarray}
and the excited states are: 
\begin{eqnarray}
    \ket{e_1} &=& \alpha \ket{a_{1g},\uparrow} - \beta \ket{a_{1g},\downarrow} \nonumber \\
    \ket{e_2} &=& \beta \ket{a_{1g},\uparrow} + \alpha \ket{a_{1g},\downarrow}.
\end{eqnarray}
where $\alpha = \cos(\theta/2)$, $\beta = \sin(\theta/2)$, and $\theta$ is the angle between the magnetic field orientation and the symmetry axis of the NiV$-$ defect. Each state’s label corresponds to its orbital and spin state. (See \cite{thiering_magneto-optical_2021} for further information on the NiV– orbital structure.) Because both excited states have allowed decay paths to multiple ground states, this system has no inherent cycling transitions (Fig. \ref{fig:NiVStrongDrive}(b)). Additionally, although significant experimental progress has been made in realizing high-Q optical photonic cavities in diamond \cite{burek_fiber-coupled_2017,guo_tunable_2021,chia_development_2022}, the narrowest linewidth cavities are still on the order of a few GHz \cite{ding_high-q_2024,stas_robust_2022},
too broad to selectively enhance the spin-conserving optical transitions, which are separated by at least a few GHz for most quantum information applications.

A strong drive is a natural choice for inducing cycling transitions in this system, given that its TDMs are of the form in Eq. 2. In Fig. \ref{fig:NiVStrongDrive}(c), we show simulation results indicating an increase in emitted photons (compared to the traditional readout method) of more than two orders of magnitude across various magnetic fields. A more misaligned field decreases the cyclicity of the bare states, but larger drive strength can compensate and allow for good readout. The strong drive effectively recreates the excited states that exist under an aligned magnetic field. 

Traditionally, there is a fundamental tradeoff between the efficiency of spin state initialization under optical pumping and the fidelity of optical readout. For example, an NiV– in an aligned magnetic field with native cycling transitions has high-fidelity state readout, but inefficient optical pumping. For an NiV$-$ in a misaligned magnetic field, optical initialization through optical pumping is efficient, but traditional single-shot state readout has low fidelity. However, our proposed strong drive scheme releases this constraint. We can efficiently optically initialize under a misaligned magnetic field, and then use a strong driving field to induce artificial cycling transitions and obtain high fidelity state readout.

The strong driving technique is extremely sensitive to the polarization of the driving field. Under ideal polarization, if the system starts in $\ket{g_2}$, then the system remains in $\ket{g_2}$ and stays dark. Under imperfect polarization, the driving field will couple $\ket{g_2}$ to the excited states, which additionally have some decay to $\ket{g_1}$, adding a source of error to the measurement. In Fig. \ref{fig:NiVStrongDrive}(d), we compare the number of photons emitted when the system starts in $\ket{g_1}$ versus in $\ket{g_2}$ for various polarization errors. The simulation is run over time $t=1e3/\gamma_0$ and with an optical Rabi rate of $\Omega_S = 100$ GHz.  The ratio is insensitive to the magnetic field orientation. The fidelity of the measurement will depend on $n_{ph,g1}$ and $n_{ph,g2}$ as well as the photon collection efficiency and the noise floor of the experimental system in question.

The NiV$-$ defect's lowest ground spin states are split by about the same energy as the excited spin states, but in systems where the excited state splitting $\omega_\textrm{ess}$ is significantly smaller than the ground states splitting $\omega_\textrm{gss}$, strong driving could be achieved in a regime where $\omega_\textrm{gss} \gg \Omega_S \gg \omega_\textrm{ess}$, which would significantly suppress this source of error because the spurious driving between $\ket{g_2}$ and the excited states would be far off-resonant.

In order to spectrally distinguish the emitted photons from the strong driving field, the measurement readout could be performed either on the phonon sideband or on a Mollow sideband, separated by $\pm \Omega_S$ from the drive frequency.

\section{Dressed excited states created by an auxiliary drive} \label{sec:auxiliary drive}

There are a few drawbacks to the strong drive method described in sections \ref{sec:strongDrive} and \ref{sec:NiV}. For one, careful polarization control and high optical powers are required. From experimental work done on the NiV$-$ \cite{morris_transition-metal_2026}, we estimate that the power needed to achieve an optical Rabi rate of $\Omega_S = 50$GHz is on the order of 10mW. In addition, a strong optical drive will only induce cycling transitions in an atomic system with a specific relationship between its TDMs. 

\begin{figure}
    \centering
    \includegraphics[width=1\linewidth]{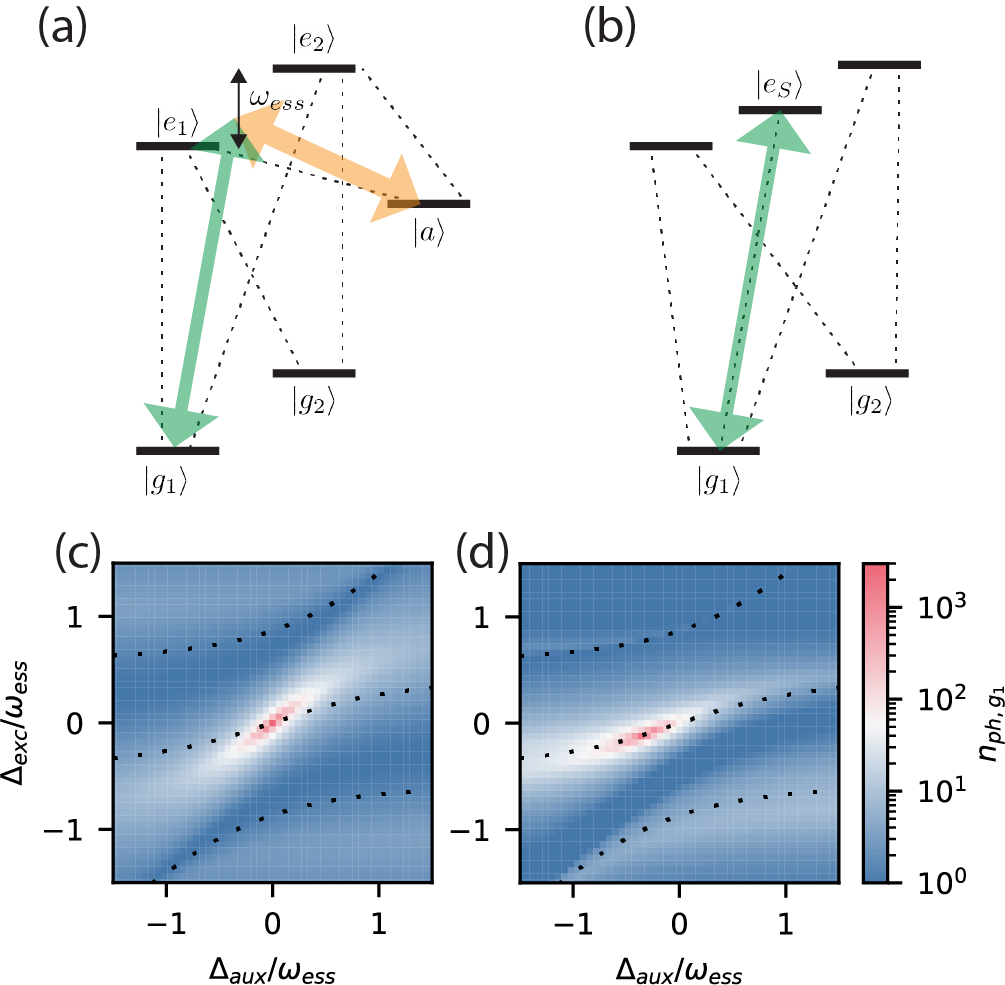}
    
    \caption{Auxiliary drive method for inducing cycling transitions. \textbf{(a)} An atomic system with no inherent cycling transitions between bare excited states $\ket{e_1}$, $\ket{e_2}$ and ground states $\ket{g_1}$, $\ket{g_2}$. Allowed transitions are shown with dotted lines. An applied auxiliary field (shown in orange) dresses the excited states by coupling them to an auxiliary state $\ket{a}$. \textbf{(b)} The same atomic system, with the excited state manifold written in the dressed basis that includes the auxiliary field. For certain power and detuning parameters of the auxiliary field, one of the dressed states ($\ket{e_S}$) does not decay to $\ket{g_2}$, yielding a cycling transition, addressed by the excitation field (shown in green). \textbf{(c,d)} Simulation of the system shown in (a,b) with TDMs given in (c) Eq. \ref{eq:balancedTDMs} and (d) Eq. \ref{eq:unbalancedTDMs}. The number of total emitted photons $n_{ph,g_1}$ over time t = 1e4/$\gamma_0$ is plotted against the detuning of the auxiliary and excitation fields. The eigenenergies of the states dressed by the auxiliary field are shown with black dotted lines. At the field detunings that yield a cycling transitioin, multiple thousands of photons are emitted with relatively low powers. For (c) the auxiliary field Rabi frequencies are $|\Omega_{e_1,a}| = |\Omega_{e_2,a}| = \omega_\textrm{ess}$, and the excitation field Rabi frequencies are $|\Omega_{e_1,g_1}| = |\Omega_{e_2,g_1}| = 0.02 \omega_\textrm{ess}$. For (d) the auxiliary field Rabi frequencies are $|\Omega_{e_1,a}| = |\Omega_{e_2,a}| = \omega_\textrm{ess}$, and the excitation field Rabi frequencies are $|\Omega_{e_1,g_1}| = 3|\Omega_{e_2,g_1}| = 0.02 \omega_\textrm{ess}$. In (c,d) the bare state decay rate from $\ket{e_1}$ to $\ket{g_1}$ is $\gamma_0 = 0.1\omega_\textrm{ess}/2\pi$.
    }
    \label{fig:AuxDriveLevels}
\end{figure}

This section describes another possible scheme for induced cycling transitions, which does not have the same drawbacks. It relies on an auxiliary field which couples the excited states to an auxiliary state $\ket{a}$ (Fig. \ref{fig:AuxDriveLevels}(a)). This creates three different dressed states that are each a superposition of $\ket{e_1}$, $\ket{e_2}$, and $\ket{a}$ (Fig. \ref{fig:AuxDriveLevels}(b)). In the strong drive scheme, the coefficients in the superposition of excited states are set by the TDMs, and the driving field must be large enough such that its Rabi frequency is larger than the energy splitting between the excited states, with larger field amplitudes being strictly better. In contrast, the auxiliary field provides an additional degree of freedom.

In order to observe an induced cycling transition the TDMs corresponding to $\ket{g_2}$ must be parallel or antiparallel. We write this as $\boldsymbol{\mu_{e_1g_2}} = \chi \boldsymbol{\mu_{g_2}}$ and $\boldsymbol{\mu_{e_2g_2}} =  \eta \boldsymbol{\mu_{g_2}}$, where $\chi$ and $\eta$ are complex coefficients. However the forms of the other TDMs are less important because the excited state superpositions can be tuned via the detuning and amplitude of this field. If a dressed excited state $\ket{e_S} = c_{e_1} \ket{e_1} + c_{e_2} \ket{e_2} + c_{a}\ket{a}$ can be created such that:
\begin{equation}
    \frac{c^*_{e_1}}{c^*_{e_2}} = -\frac{\eta}{\chi},
\end{equation}
then the excitation drive frequency can be chosen to selectively couple $\ket{g_1}$ and $\ket{e_S}$, forming a cycling transition.

In Fig. \ref{fig:AuxDriveLevels}(c), we simulate the behavior of the atomic system shown in Fig. \ref{fig:AuxDriveLevels}(a,b), assuming the following TDMs:
\begin{eqnarray}\label{eq:balancedTDMs}
\boldsymbol{\mu_{e_1g_1}} &=&\boldsymbol{\mu_{g_1}} \nonumber\\
\boldsymbol{\mu_{e_2g_1}} &=&  -\boldsymbol{\mu_{g_1}} \nonumber\\
\boldsymbol{\mu_{e_1g_2}} &=& \boldsymbol{\mu_{g_2}} \nonumber\\
\boldsymbol{\mu_{e_2g_2}} &=&  \boldsymbol{\mu_{g_2}} \nonumber\\
\boldsymbol{\mu_{e_1a}} &=& \boldsymbol{\mu_{a}} \nonumber\\
\boldsymbol{\mu_{e_2a}} &=&  \boldsymbol{\mu_{a}} 
\end{eqnarray}
where $|\boldsymbol{\mu_{g_1}}| = |\boldsymbol{\mu_{g_2}}|$. Other than the presence of the auxiliary state, these TDMs are equivalent to those in the atomic system shown by the green line ($b=0.5$) in Fig. \ref{fig:strongDriveComparison}(a).

We sweep over the detunings of the auxiliary drive and the excitation drive, defined as:
\begin{eqnarray}
\Delta_\textrm{aux}&=&\omega_\textrm{aux}-((\omega_{e_1}+\omega_{e_2})/2 - \omega_{a})\nonumber\\
\Delta_\textrm{exc}&=&\omega_\textrm{exc}-((\omega_{e_1}+\omega_{e_2})/2 - \omega_{g_1})
\end{eqnarray}
where $\omega_\textrm{aux}$ ($\omega_\textrm{exc}$) is the frequency of the auxiliary (excitation) drive and $\omega_i$ is the frequency of state $i$. As the auxiliary drive frequency varies, the characteristics of the dressed excited states change, and we probe these changing excited states via the excitation drive. When $\Delta_\textrm{aux} = 0$, the transition between $\ket{g_1}$ and the dressed state $\ket{e_S}$ becomes cycling, and when the excitation drive is resonant with this transition ($\Delta_\textrm{exc} = 0$) we find that over time t=1e4/$\gamma_0$, there are $n_{ph,g_1}$=3040 emitted photons. The simulation is stopped at this point to reduce computation time, but the system continues to emit photons.

In the absence of the auxiliary drive, this atomic system would require optical Rabi frequencies of $\Omega_S > 50\omega_\textrm{ess}$  in order to achieve an equivalent value of $n_{ph,g_1}$ (Fig. 2(a)), but the auxiliary drive technique simulated here has much lower driving powers, with auxiliary field Rabi frequencies of $|\Omega_{e_1,a}| = |\Omega_{e_2,a}| = \omega_\textrm{ess}$, and excitation field Rabi frequencies of $|\Omega_{e_1,g_1}| = |\Omega_{e_2,g_1}| = 0.02\omega_\textrm{ess}$.

In Fig. \ref{fig:AuxDriveLevels}(d), we simulate the behavior of an atomic system with a different set of TDMs:
\begin{eqnarray}\label{eq:unbalancedTDMs}
\boldsymbol{\mu_{e_1g_1}} &=&\boldsymbol{\mu_{g_1}} \nonumber\\
\boldsymbol{\mu_{e_2g_1}} &=& -0.333\boldsymbol{\mu_{g_1}} \nonumber\\
\boldsymbol{\mu_{e_1g_2}} &=& 0.333\boldsymbol{\mu_{g_2}} \nonumber\\
\boldsymbol{\mu_{e_2g_2}} &=& 0.5\boldsymbol{\mu_{g_2}} \nonumber\\
\boldsymbol{\mu_{e_1a}} &=& \boldsymbol{\mu_{a}}\nonumber\\
\boldsymbol{\mu_{e_2a}} &=&  \boldsymbol{\mu_{a}} 
\end{eqnarray}
Other than the presence of the auxiliary state, these atomic system parameters are illustrated with a dashed blue line in Fig. \ref{fig:strongDriveComparison}(b). Under a single strong drive, a cycling transition is not achievable and $n_{ph,g_1}$ saturates at a few tens of photons. However, by applying an auxiliary drive with a Rabi frequency of $|\Omega_{e_1,a}| = |\Omega_{e_2,a}| = \omega_\textrm{ess}$ and a detuning of $\Delta_\textrm{aux} = -0.31\omega_\textrm{ess}$, a dressed excited state is generated which forms a cycling transition with $\ket{g_1}$. With excitation Rabi frequencies of $|\Omega_{e_1,g_1}| = 3|\Omega_{e_2,g_1}| = 0.02\omega_\textrm{ess}$ and detuning of $\Delta_\textrm{exc} = -0.10\omega_\textrm{ess}$, this cycling transition produces 2810 photons over the simulation time t=1e4/$\gamma_0$ (Fig. \ref{fig:AuxDriveLevels}(d)).

\section{Discussion}

We have proposed and analyzed two methods for enhancing the cyclicity of optical transitions in atomic systems by exploiting the coherent interference of decay paths. By leveraging quantum interference in the decay process, they circumvent the need for narrow optical cavities or highly selective transitions, offering a scalable route to improved optical readout fidelity in quantum technologies. 

The technique discussed in Sections \ref{sec:strongDrive} and \ref{sec:NiV} relies on the ability to drive solid-state qubits with large optical powers, a capability which is ever-increasing as active materials engineering efforts improve crystal quality and reduce spurious optical absorption \cite{zhang_enhanced_2024,inaba_er-doped_2024,kavatamane_reversing_2026}. There are also further opportunities to investigate interference across three or more decay paths. Because three non-parallel vectors can sum to zero, these paths can destructively interfere even if the associated TDMs are not parallel. These paths have the potential to expand the number of systems in which effective cycling transitions can be engineered, across a wide range of atomic and solid-state systems.

\begin{acknowledgments}
The authors acknowledge Hoang Le, Aaron Day, and Eric Cornell for useful discussion.
This research was supported by the National Science Foundation under grant numbers 606947 and 2014Y0A, the Air Force Office of Scientific Research under award number 1563927, and Amazon Web Services under award number A60290. B.P. acknowledges financial support from the US Department of Energy, Office of Science, Basic Energy Sciences, Materials Sciences and Engineering Division through Argonne National Laboratory under contract number DE-AC02-06 CH11357. A.G.\ acknowledges funding from the European Commission for the SPINUS (Grant No.\ 101135699) and QuSPARC (Grant No.\ 101186889) projects. G.T.\ acknowledges the STARTING Grant No.~150113 from the National Office of Research, Development and Innovation of Hungary (NKFIH) and the support from the J\'anos Bolyai Research Scholarship of the Hungarian Academy of Sciences.

\end{acknowledgments}

\appendix

\section{Coherent interference of decay}

A full master equation description of an atomic system includes collapse operators:
\begin{equation}
    \frac{\partial}{\partial t} \rho = -\frac{i}{\hbar} [H,\rho] - \frac{1}{2} \sum_{j,k} \Gamma_{jk} (C_j^+ C_k^- \rho + \rho C_j^+ C_k^- - 2 C_k^- \rho C_j^+)
\end{equation}
where $j$ and $k$ label the different transitions between states. The cross damping rate for two different transitions ($j\neq k$) is given by $\Gamma_{jk} = \sqrt{\Gamma_j \Gamma_k} \cos \theta_{jk}$ where $\theta_{jk}$ is the angle between the $j$ and $k$ dipole moment polarizations. When non-zero, the cross-damping terms interfere with the diagonal damping terms and can result in cancellation of the decay path from certain superposition states. Thus, the polarization of the transition dipole moments (TDMs) of the given decay paths is of utmost importance.

This can more intuitively be understood as arising from the effective TDMs between a superposition state and a ground state. For example, assuming that $\boldsymbol{\mu_{e_1g_2}} = \bra{e_1} \boldsymbol{d}\ket{g_2} = \beta_1^* \boldsymbol{\mu_-}$ and $\boldsymbol{\mu_{e_2g_2}} = \bra{e_2} \boldsymbol{d}\ket{g_2}= \alpha_1^* \boldsymbol{\mu_-}$ (where $\boldsymbol{d}$ is the dipole operator), then the superposition state $\ket{e_S} \sim \alpha_1 \ket{e_1} - \beta_1 \ket{e_2}$ has no decay path into state $\ket{g_2}$ because $\boldsymbol{\mu_{e_Sg_2}} = \bra{e_S} \boldsymbol{d}\ket{g_2} = 0$.

It is additionally important to consider time-dependence of operators and states when determining whether a system displays cancellation of decay. When atomic states are coupled by an external oscillating field, the time-dependence can be canonically removed from the Hamiltonian by taking a transformation in which states are redefined in the form $\ket{e_1} = \ket{\tilde e_1} e^{i\omega t}$. This allows us to derive time-independent dressed states from the Hamiltonian. However, this transformation can in some cases introduce a time-dependence in the collapse operators which prevents the effective increase in cyclicity.

\begin{figure}
    \centering
    \includegraphics[width=1\linewidth]{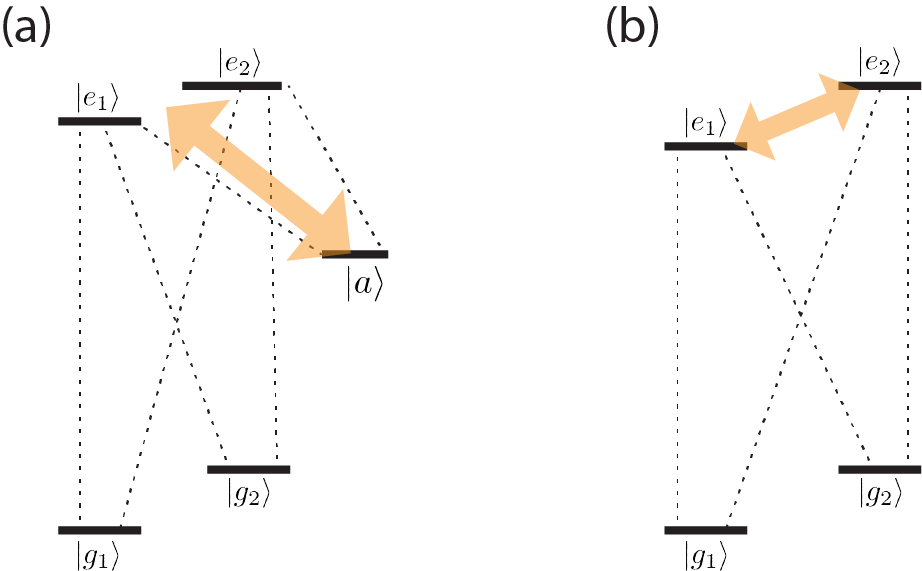}
    \caption{\textbf{(a)} A system in which coherent interference of decay could be observed. \textbf{(b)} A system in which coherent interference of decay could not be observed due to time-dependence in the collapse operator terms.
    }
    \label{fig:directDrivingNoGo}
\end{figure}

Consider the two examples of atomic systems in Fig. \ref{fig:directDrivingNoGo}. In Fig \ref{fig:directDrivingNoGo}(a), state $\ket{a}$ is simultaneously coupled to states $\ket{e_1}$ and $\ket{e_2}$. This is described by adding an identical time-dependence to states $\ket{e_1}$ and $\ket{e_2}$, which cancels out of the master-equation decay terms. Equivalently, a time-dependence could be added to state $\ket{a}$, but this likewise has no effect on the relevant decay terms because state $\ket{a}$ does not decay to the ground state manifold.

On the other hand, in the system shown in Fig. \ref{fig:directDrivingNoGo}(b) a field directly couples states $\ket{e_1}$ and $\ket{e_2}$. In order to eliminate the time-dependence from the Hamiltonian and define its stationary states, states $\ket{e_1}$ and $\ket{e_2}$ must have different time-dependences, which do not cancel out of the master-equation decay terms. This prevents a measurable coherent interference of decay from occurring.

In general, in order for decay from different states to interfere, they must be simultaneously coupled by the same fields to the same states.

\begin{figure}
    \centering
    \includegraphics[width=1\linewidth]{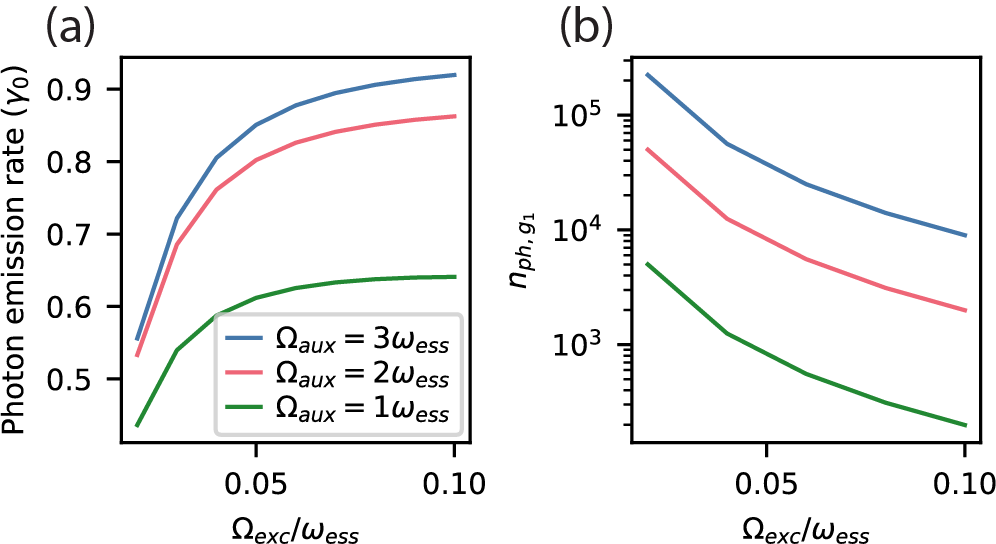}
    \caption{Tradeoff between photon emission rate and the total number of emitted photons for the system described in Eq. \ref{eq:balancedTDMs} and simulated in Fig. \ref{fig:AuxDriveLevels}(c). \textbf{(a)} The photon emission rate in units of the bare excited state decay rate $\gamma_0$. As the excitation power increases, photons are emitted at a faster rate. \textbf{(b)} The total emitted photons $n_{ph,g_1}$. Larger excitation powers lead to greater off-resonant excitation of additional excited states which decay to $\ket{g_2}$, reducing the total number of emitted photons. For either figure of merit, it is beneficial to have a larger auxiliary field amplitude $\Omega_\textrm{aux}$.
    }
    \label{fig:auxLevelsPowerTradeoff}
\end{figure}

\section{Photon-counting figure of merit} \label{app:Photon counting FOM}
We calculate the number of emitted photons from the evolution of the system by using the terms in the master equation that directly correspond to photon emission. The rate of photon emission $r_{ph}$ is:
\begin{equation}
    r_{ph} = \ \dot \rho_{g_1g_1}^{dec} + \dot \rho_{g_2g_2}^{dec}
\end{equation}
where $\dot \rho_{ij}^{dec}$ is the change in the $ij$-component of the density matrix due to decay operators as opposed to coherent evolution. We write in terms of the collapse operators:
\begin{eqnarray}
r_{ph}=\bra{g_1}\Bigl( - \frac{1}{2} \sum_{j,k} &\Gamma_{jk}& (C_j^+ C_k^- \rho + \rho C_j^+ C_k^- \nonumber\\
&&- 2 C_k^- \rho C_j^+) \Bigr) \ket{g_1} \nonumber\\ 
+\bra{g_2}\Bigl( - \frac{1}{2} \sum_{j,k} &\Gamma_{jk}& (C_j^+ C_k^- \rho + \rho C_j^+ C_k^- \nonumber\\
&&- 2 C_k^- \rho C_j^+) \Bigr) \ket{g_2}.
\end{eqnarray}
Since the collapse operators are all of the form $C^-=\ket{g}\bra{e}$, we can simplify:
\begin{equation}
    r_{ph} =   \sum_{j,k} \Gamma_{jk} \Bigl( \bra{g_1}C_k^- \rho C_j^+ \ket{g_1} + \bra{g_2}C_k^- \rho C_j^+\ket{g_2}  \Bigr).
\end{equation}
We can add the terms
\begin{equation}
    \bra{e_1}C_k^- \rho C_j^+ \ket{e_1} + \bra{e_2}C_k^- \rho C_j^+\ket{e_2}
\end{equation}
inside the sum since they are both zero, rewrite the expression in terms of the trace, and
rearrange suggestively:
\begin{equation}
    r_{ph} =   \sum_{j,k} \Gamma_{jk} Tr[C_k^- \rho C_j^+] = \sum_{j,k} \Gamma_{jk} \langle C_j^+ C_k^-\rangle 
\end{equation}
All results in this paper that state a number of photons $n_{ph}$ emitted from a system are obtained using this expression. The NiV$-$ simulations in section \ref{sec:NiV} include the orbital decay from the upper orbital ground states (not pictured in Fig. \ref{fig:NiVStrongDrive}) to the lower orbital ground states with a transition frequency of 100s of GHz, but these events are not included as measurable emitted photons.

\section{Excitation strength tradeoff in the auxiliary method}

Consider the system illustrated in Fig. \ref{fig:AuxDriveLevels}(b). In the ideal case the excitation field exclusively excites $\ket{g_1}$ to $\ket{e_S}$. However, given finite system parameters, the excitation field additionally off-resonantly excites $\ket{g_1}$ to the other dressed states, which decay to $\ket{g_2}$, reducing the cyclicity. This effect scales with $\Omega_\textrm{exc}/(\Delta_\textrm{dressed})^2$, where $\Omega_\textrm{exc}$ is the optical Rabi frequency of the excitation field and $\Delta_\textrm{dressed}$ is the detuning between dressed excited states. We can suppress this effect by increasing $\Omega_\textrm{aux}$, which increases the splitting between the dressed states. We can alternatively reduce our excitation power, but this reduces the rate at which photons are emitted. In Fig. \ref{fig:auxLevelsPowerTradeoff} we simulate this tradeoff between photon rate and total expected number of photons emitted.

\section{Auxiliary drive method atomic structures}
Here we describe the necessary attributes of an auxiliary state $\ket{a}$ for use in producing cycling transitions as described in Section \ref{sec:auxiliary drive}. In order to produce a stable superposition, the decay and decoherence between $\ket{e_1}$, $\ket{e_2}$, $\ket{a}$ must be negligible. This may be the case when the energy difference between $\ket{a}$ and the excited states is much smaller than the energy difference between the excited states and the ground states. In addition, although a decay channel from $\ket{a}$ to $\ket{g_1}$ does not have a significant effect on the system's cyclicity, a decay channel from $\ket{a}$ to $\ket{g_2}$ suppresses the cyclicity of the system.

\bibliography{bib_v5}

\end{document}